\documentclass[twocolumn,final,fleqn]{svjour2}
\usepackage{pslatex}
\usepackage{latexsym}
\usepackage{amsmath,amssymb}
\usepackage{graphicx}
\usepackage{dcolumn}
\usepackage{bm}
\usepackage[numbers]{natbib}
\journalname{Granular Matter}

\begin{document}
\title{Pressure and Motion of Dry Sand -- Translation of 
Hagen's Paper from 1852}
\author{Brian P. Tighe \and Matthias Sperl}
\institute{Duke University, Department of Physics, Box 90305,\\
Durham, NC 27708, USA,\\\email{msperl@duke.edu}
}
\date{\today}

\maketitle 
\begin{abstract}
In a remarkable paper from 1852, Gotthilf Heinrich Ludwig Hagen measured 
and explained two fundamental aspects of granular matter: The first effect 
is the saturation of pressure with depth in a static granular system 
confined by silo walls -- generally known as the \textit{Janssen effect}. 
The second part of his paper describes the dynamics observed during the 
flow out of the container -- today often called the \textit{Beverloo law} 
-- and forms the foundation of the \textit{hourglass theory}. The 
following is a translation of the original German paper from 1852.
\end{abstract}

\section{Introduction}

Gotthilf Heinrich Ludwig Hagen is most renowned for his contributions to 
the study on laminar flow in pipes; his measurements published in 1839 
studied what is now well-known as the (Hagen-)Poiseuille law 
\cite{Hagen1839,Sutera1993}. Less well-known is Hagen's work on granular 
systems. While Janssen, with his 1895 paper, typically receives credit for 
the saturation effect in granular silos \cite{janssen95,Sperl2006b}, it 
was Hagen in his paper \textit{\"Uber den Druck und die Bewegung des 
trocknen Sandes} \cite{Hagen1852} who measured this effect earlier -- but 
also not for the first time, cf. \cite{Huber1829} -- and offered a first 
model that provided a qualitative understanding of the effect. Hagen 
proposes a quadratic law (with some cutoff) for the pressure instead of 
the exponential form put forward by Janssen more than 40 years later 
\cite{Sperl2006b,nedderman92}.

In addition to the discussion of a static pile of sand, Hagen examines in 
considerable detail the flow through an opening of the container, and 
discovers that the rate of discharge is proportional to the diameter of 
the opening raised to the power $5/2$. This result is elegantly derived 
from dimensional analysis \cite[ch 10.2]{nedderman92}; but it fits the 
data best if instead of the real diameter some effective diameter is used 
-- the resulting law is known as the Beverloo correlation 
\cite{Beverloo1961}. Hagen finds an effective opening diameter that is 
smaller by twice the particle diameter, which is consistent with more 
recent measurements.

Hagen's work lays out the basis of the so-called hourglass theory where 
the flow of granular material is found to be independent of the filling 
height in the container, thus allowing the measure of time with an 
hourglass \cite[sec 10.4]{nedderman92}. Later work both confirms and 
extends Hagen's early analysis \cite{Wieghardt1975}. More recently, a 
lot of work has been devoted to understanding the fundamental differences 
between fluid and granular flows \cite{Rajchenbach2000}. While on the 
level of the individual grains the probability of arching at the opening 
is a matter of current investigations \cite{To2001}, Hagen provides a 
successful route to a continuum description of the flow by the rescaling 
of the diameter of the opening.

\section{Gotthilf Heinrich Ludwig Hagen, \textit{Pressure and Motion of 
Dry Sand}, 19th January 1852}

Consider a container with a horizontal bottom that includes a circular 
opening of radius $r$. In this opening is placed a disk which is easily 
movable but seals tightly; on top of it is an extended filling of sand up 
to a height $h$. As a result, there is a pressure exerted on the disk 
created by the weight of the cylinder of sand above it less the friction 
which is experienced by that cylinder from the sand surrounding it. The 
friction is proportional to the horizontal pressure, or the square of the 
height. Let $l$ denote a friction dependent constant and $\gamma$ be the 
weight of a unit volume of sand, then the pressure against the disk equals

\begin{equation*}
r^2\pi\gamma h - 2 r \pi\gamma l h^2.
\end{equation*}

For a growing $h$, this expression will increase in the beginning, reach a 
maximum, and subsequently decrease afterwards; it will become not only 
zero but even negative. However, the sand cylinder is not rigidly 
connected throughout, and therefore the axial pressure, which its lower 
part exerts on the bottom disk, cannot be compensated by the strong 
friction acting on the cylinder as a whole. Hence, the pressure on the 
disk in fact remains unchanged for fillings higher than that height at 
which the pressure on the disk reaches its maximum value.

For the case of pressures below the maximum, we seek to represent that 
pressure by the weight of a free-standing body of sand on the disk. The 
body is bounded by the surface of the filling. It is a conoid that is 
formed by the rotation of a parabola around its axis. Here the parameter 
of the parabola is $4 r l$, while the height of the paraboloid is 
$r/(4l)$. The latter body joins the circumference of the opening.

To compare these results with the real phenomenon, I created openings of 
radii of 0.3791 inch\footnote{trans. note: The German \textit{Zoll} is 
translated as \textit{inch}; however, it might deviate slightly from the 
value generally used today; most likely $1\ {\rm Zoll} = 26.15{\rm mm}$ 
\cite{Trapp1996}.} and 0.7271 inch, respectively, in two brass plates used 
as the the bottom of the sand-filled container; these openings were closed 
with suitable disks that were supported from underneath by hooks that were 
connected to one arm of a balance, while the other arm carried the 
counterweight. To reduce the counterweight to the pressure on the disk 
slowly and without concussion, sand was allowed to discharge through a 
small hole in the bottom of the plate. The constant error of this method 
could be found easily by measurement of the excess weight of the disk and 
the hook compared to the plate both by the discharge of sand and by direct 
weighing.

While repeating measurements of the pressure due to the the sand against 
the disks multiple times, considerable deviations among the measurements 
were observed; this was apparently due to different ways of settling. When 
the settling was as loose as possible, the weight of a cubic inch of the 
sand, a crude ferrous grit, was 2.9~Loth\footnote{trans. note: The German 
\textit{Loth} (or \textit{Lot}) is a unit of mass, $1~{\rm Loth} = 
1/32~{\rm Handelspfund(Berlin)}\approx 14.6{\rm g}$ \cite{Trapp1996}.}. 
However, the weight increased to 3~Loth when there was modest agitation 
during filling and rose to 3.25~Loth as soon as a fairly compact settling 
was generated by severe agitation or by pushing a wire into the container. 
In so doing, the friction increased by even more than the specific weight. 
Hence, the pressure against the disk became remarkably smaller for the 
more compact settling.

With the larger disk the pressure maximum was reached at a filling height 
of around 1~inch: for larger height the pressure decreased somewhat, since 
despite great care the sand settled in a slightly denser state. For the 
loosest fillings I found $l = 0.154$ to 0.175. In contrast, the deviations 
were less when the sand was dropped from a height of several inches in a 
narrow stream, and when the sand flow was cone-shaped: the value for $l$ 
was limited to between 0.21 and 0.22.

The influence of different settling states was also clearly noticeable as 
the sand discharged through the openings of various radii. When the 
settled filling was somewhat denser, there was less sand flowing within a 
second. As the discharge duration increased, the sand became agitated, 
particularly in the vicinity of the opening, resulting in an increased 
sensitivity, which in turn led to a diminished flow. Incidentally, the 
height of the filling had no influence, as was already recognized by 
Huber-Burnand some time ago \cite{Huber1829}\footnote{trans. note: The 
reference was added for the translation. There are no references given in 
the original.}. To reduce the mentioned irregularities as much as 
possible, I limited the duration of each observation to 30 to 200 seconds 
and, additionally, tried to make the fillings rather uniform. To this end, 
the container was placed in a metal-sheet cylinder with a sieve-like 
bottom, filled with sand, and lifted slowly thereafter; in this way the 
sand poured into the container in several hundred thin streams, each from 
a very small height.

The openings at the bottom of the container, with their sharp edges always 
on the upper face, had the following radii: 

\begin{tabular}{lrlc}
opening & 1 & 0.1677 & inch \\
opening & 2 & 0.1203 & inch \\
opening & 3 & 0.0986 & inch \\
opening & 4 & 0.0807 & inch \\
opening & 5 & 0.0549 & inch \\
opening & 6 & 0.0377 & inch 
\end{tabular}

The amount of outflowing sand per second, averaged over six observations 
each, was:

\begin{tabular}{lrlc}
for the opening & 1& 1.8995 & Loth \\
for the opening & 2& 0.7596 & Loth \\
for the opening & 3& 0.4330 & Loth \\
for the opening & 4& 0.2481 & Loth \\
for the opening & 5& 0.08242 & Loth \\
for the opening & 6& 0.02676 & Loth \\
\end{tabular}

Comparing these weights with the radii of the openings, the former seem to 
be approximately proportional to the third power of the latter. An attempt 
to represent them in the form \[ m = k r^3, \] however, did not produce a 
satisfactory result; the remaining errors were rather significant and very 
regular, so they could not be interpreted as observational errors. In 
contrast, if the radius of the opening was reduced by some length, a 
nearly constant ratio was observed between the mass of sand and the 
$2.5^{\mathrm th}$ power of the reduced radius. The reduction of the 
radius is justified by noting that the granules that touch the edge of the 
opening while falling lose their speed partially or completely and even 
disturb the motion of the neighboring granules when bouncing off. From 
repeated measurements it was found that, on average, 9 granules of sand 
constitute the length of a Rhineland line\footnote{trans. note: The German 
\textit{Rheinl\"andische Linie} is a unit of length, typically 1/12 of a 
Zoll, hence $1\ {\rm line} = 2.18{\rm mm}$ \cite{Trapp1996}.} hence the 
diameter of a single one equals 0.0093 inches.

Thus I compared the masses of the sand with the expression
\[
m = k (r-x)^\frac{5}{2}
\]
and found after introducing approximate values for $x$ according to the 
method of least squares from all six observations 
\[\begin{array}{lcl}
k &=& 189.07 \\
x &=& 0.00968.
\end{array}\]
Using these values to calculate $m$, the remaining errors are
\begin{tabular}{lrl}
for & 1& -0.0181 \\
for & 2& +0.0094 \\
for & 3& +0.0135 \\
for & 4& +0.0075 \\
for & 5& -0.00003 \\
for & 6& -0.00142 
\end{tabular}

The first observation agrees the least with the fitted values, but is also 
in itself less accurate than the other observations for two reasons: for 
one part, its duration was the shortest, and for the other part, the sand 
was poured out vehemently and became heavily agitated, and the container 
was almost completely emptied. Therefore, I tried to determine the values 
of the two constants independently of the first observation, using only 
the last five. This gives
\[m = 181.57(r-0.00893)^\frac{5}{2}.\]
The deviations from the observed weights were thereupon
\begin{tabular}{lrl}
for & 1& -0.0712 \\
for & 2& -0.0085 \\
for & 3& +0.0050 \\
for & 4& +0.0039 \\
for & 5& +0.00002 \\
for & 6& -0.00 078
\end{tabular}

The radius reduction $x$ for the discharge opening is found to be close to 
the diameter of a grain of sand. From the constant $k$ can be determined 
the average distance from the opening at which the sand begins its free 
fall. Assuming that the sand forms a compact mass until reaching the 
opening, the amount of sand discharged per second is
\[m = 2\rho^2\pi\gamma\sqrt{g h }\]
where $h$ designates the mentioned height of fall and $\rho$ the effective
opening. On the other hand, the observations yield
\[m = 181.57 \cdot \rho^\frac{5}{2}.\]
Equating both expressions and setting $\gamma$ to 2.93 as obtained from the 
average loose packings, one finds
\[h = 0.5185\cdot\rho.\]

If, in contrast, one assumes that for each unit of time a layer of sand of 
the same vertical height separates from the whole inner area of the 
paraboloid mentioned above in a free fall, then one can easily find the 
average velocity of this layer while passing the opening, and from the 
latter the average height of fall of the entire mass. This height is
\[h = \frac{r}{9l}.\]
But from the data for loose packings it was found that
\[l = 0.16\]
and hence
\[h=0.6944\cdot r.\]
If instead one introduces the value
\[l = 0.225,\]
which is valid for packings where sand is flowing sideways,
which really happens during the discharge, then  
\[h = 0.4938\cdot r.\]
The result derived from the observations is in between the two when $r$ is 
interchanged with $\rho$. This interchange is necessary because the sand 
only hits the edge of the opening when in motion; in contrast, when at 
rest all the sand grains encountering the movable disk also load it. This 
confirms the assumption from above that the free fall of the sand starts 
on the surface of the paraboloid; it also explains that the amount of sand 
flowing through the opening is proportional to the power $\frac{5}{2}$ of 
the effective radius of the opening.

Finally, some comments should be made about the motion of the sand 
during discharge.

In the four inch wide and ten inch tall container, above the outlet, the 
entire surface of the sand packing subsided uniformly in the beginning. 
Only gradually did a dip form vertically above the outlet. The dip grew 
continually, and above its sides sand fell down. Concurrently, at the rim 
of the container a ring-shaped, almost horizontal surface remained. This 
surface also subsided, but without the granules of sand experiencing 
strong sideways motion. The flat ring gradually assumed a smaller width 
and disappeared completely when the funnel-shaped dip reached the outlet. 
From this it follows that the sand flows not only vertically towards the 
outlet but also along concentric inclined trajectories, and that the 
motion extends up to a slope, which a free surface of sand can exhibit.

The motion in the inner part of the sand mass revealed itself very 
explicitly when I filled sand in a container having side walls made from
a glass panel. Since this glass panel touched the outlet, one could 
follow the motion of single grains of sand down to the opening. The 
strongest flow formed vertically above the outlet; in fact the sand 
granules approached it with increasing yet moderate speed until, directly 
above, they were accelerated in a way that they could no longer be seen. 
Nevertheless, the sand also flowed inwards from the side of the outlet, but 
this motion was interrupted frequently and only occurred periodically, 
presumably due to friction at the glass.

Underneath the opening, the stream of sand was not nearly as sharply 
bounded as a water-jet; rather, it was surrounded by single granules that 
from time to time departed as far as several lines\footnote{trans. note: 
\textit{Linien}, plural as in \textit{Rheinl\"andische Linien}.}. Because 
of this the streams flowing from the larger outlets showed a significant 
reduction in their diameters, extending about 2~inches deep. In addition, 
the measurement showed that even immediately under the disk the stream is 
already much weaker than at the outlet. The outlet was 0.335~inch in 
diameter while the stream was only 0.29~inch at a distance of 
$1\frac{1}{2}$ lines, and contracted to 0.27~inch at greater depth. The 
reason for this effect is not the effective reduction of the opening 
mentioned above, since this would only explain the weakening of the stream 
by about 2\% of an inch; instead, the sand flowing sideways continues its 
motion towards the axis even after having passed the outlet, and the 
granules hitting the rim of the opening are also reflected towards the 
axis.

The stream of sand hence experiences a contraction similar to the stream 
of a liquid; and when one compares the diameter of the opening with the 
smallest diameter of the stream, the ratio appears as 
\[1:0.806\]
or for the ratio of cross sections
\[1:0.650\,.\]
This agrees closely with the known contraction ratios for liquid streams 
leaving openings in thin walls.

\begin{acknowledgements}
This work is supported by NSF-DMT0137119, DMS0244492, and DFG~SP~714/3-1.
\end{acknowledgements}


\bibliographystyle{apsrev}

\end{document}